\documentstyle[pra,epsf,aps,twocolumn]{revtex}
\begin{document}
\title{Effective one-body approach to the relativistic two-body problem}
\author{Plamen P.~Fiziev \thanks{ E-mail:\,\, fiziev@phys.uni-sofia.bg}}
\address{Department of Theoretical Physics,
Sofia University, 5 James Bourchier Boulevard, BG-1164, Sofia,
Bulgaria}
\author{Ivan T.~Todorov \thanks{ E-mail:\,\, todorov@inrne.bas.bg}}
\address{Institute of Nuclear Research and Nuclear Energy, Tsarigradsko Chaussee 72,
 BG-1784 Sofia, Bulgaria}
\maketitle
\begin{abstract}
The relativistic 2-body problem, much like the non-relativistic
one, is reduced to describing the motion of an effective particle
in an external field. The concept of a relativistic reduced mass
and effective particle energy, introduced some 30 years ago to
compute relativistic corrections to the Balmer formula in quantum
electrodynamics, is shown to work equally well for classical
electromagnetic and gravitational interaction. The results for the
gravitational 2-body problem have more than academic interest
since they apply to the study of binary pulsars that provide
precision tests for general relativity. They are compared with
recent results derived by other methods.

\noindent{PACS number(s): 03.30.+p, 04.25.-g, 95.30.Sf, 97.80.Fk}

\end{abstract}

%%%%%%%%%%%%%%%%%%%%%%%%%%%%%%%%%%%%%%%%%%%%%%%%%%%%%%%%%%%%%%%%%%%
%\draft
\sloppy
%\scrollmode
%%%%%%%%%%%%%%%%%%%%%%%%%%%%%%%%%%%%%%%%%%%%%%%%%%%%%%%%%%%%%%%%%%%%

\section{Introduction}

The notion of an effective relativistic particle describing the
relative motion of a 2-body system was first introduced in the
context of the quasipotential approach to the quantum field
theoretical eikonal approximation \cite{T71} and bound state
problem in quantum electrodynamics \cite{T73}, \cite{RTA75}.
For a
survey of subsequent developments - see \cite{MP}. Only later a
general classical mechanical formulation of the relativistic
2-body problem was given \cite{T76} within Dirac's constraint
Hamiltonian approach \cite{D49} (for a review - see \cite{T82}).

The central concept of a {\em relativistic reduced mass} is
derived in this early work by observing that the total mass $M$ of
a 2-particle system should be substituted by its total
centre-of-mass (CM) energy:
\begin{eqnarray}
{w \over {c^2}} = M +{1 \over {c^2}} E\,\,\, \left(\sim M
\,\,\,\,\hbox{for}\,\,\,\, {\frac {|E|} {Mc^2}} \ll 1 \right),
\label{w}
 \end{eqnarray}
$$ M=m_1+m_2.\hskip 5.4truecm$$

This suggests using an energy dependent expression
\begin{eqnarray}
m_w={{m_1m_2c^2}\over w}\,\,\,\left(\rightarrow \mu =
{{m_1m_2}\over M}  \,\,\,\,\hbox{for}\,\,\,\,{w \over
{c^2}}\rightarrow M\right),
\label{mw}
 \end{eqnarray}
for the relativistic generalization of the reduced mass $\mu$.

Furthermore, if we determine the off-shell momentum  square
$b^2(w^2)$ for a pair of free particles as the solution for ${\bf
p}^2$ of the equation
\begin{eqnarray}
{w \over c}= \sqrt{m_1^2 c^2 + {\bf p}^2} +\sqrt{m_2^2 c^2 + {\bf
p}^2}\hskip 1.truecm \Rightarrow \hskip 1.1truecm \nonumber \\
{\bf p}^2\!=\!b^2(w^2)\!=\!{{w^4 \!-\!2(m_1^2\!+\!m_2^2)c^4 w^2
\!+\!(m_1^2\!-\!m_2^2)^2c^8}\over {4 w^2 c^2}}
\label{eq13}
 \end{eqnarray}
we find for the effective particle CM energy
\begin{eqnarray}
E_w=c\sqrt{m_w^2 c^2 + b^2(w^2)} = {{w^2 -m_1^2 c^4 -m_2^2
c^4}\over {2w}}.
\label{Ew}
 \end{eqnarray}

The interest in this notion of an effective particle was revived
recently in \cite{BD99} where a modified version of it was applied
to the general relativistic 2-body dynamics and  the relevance
(and relative simplicity) of the dimensionless counterpart of the
effective particle energy
\begin{eqnarray}
\epsilon={E_w \over {m_w c^2}} = {{w^2 -m_1^2 c^4 -m_2^2 c^4}\over
{2 m_1 m_2 c^4}}
\label{epsilon}
 \end{eqnarray}
(which only makes sense for positive mass particles) was pointed
out. On the other hand, the authors of \cite{BD99} preferred to
work with the non-relativistic reduced mass $\mu$ rather than with
the energy dependent quantity $m_w$.

The present paper was motivated by our wish to demonstrate the
advantage of the original notion of relativistic effective
particle, cited in the beginning, in both the (classical)
electromagnetic and gravitational 2-body problem.

We begin in Sec.2, by recalling the constraint Hamiltonian
approach to a relativistic particle system. A new justification is
provided on the way for formula (\ref{mw}) for the energy
dependent reduced mass identified as the coefficient to the
relative velocity in the expression for the effective particle
3-momentum in the CM frame.

Sec.3 is devoted to the electromagnetic interaction of two
oppositely charged particles.

The general relativistic gravitational two-body problem (which
continues to attract attention - see, e.g. \cite{CE77},
\cite{DS88}, \cite{OK89}, \cite{SW93}, \cite{DJS00}) is addressed
in Sec.4. We compute the perihelion shift as well as the
parameters of the last stable orbit in the gravitational case and
compare with earlier results.

In both cases we neglect the retardation effect: it is known not
to contribute to the first post-Newtonian approximation (see
\cite{LL}) for which our results agree with previous calculations.
The effects of the relativistic kinematics, on the other hand, are
computed exactly.

The possibility to take into account the finite velocity of
propagation of interactions starting with the second
post-Newtonian approximation within the effective particle
approach of this paper is discussed in the concluding Sec.5.

\section{Velocity space formulation of the constraint Hamiltonian
approach to the relativistic 2-body problem}

The {\em mass-shell constraint} for a free relativistic particle
can be interpreted as a {\em "Lorentz invariant Hamiltonian"}:
\begin{eqnarray}
H={1\over {2\lambda}}\left( m^2 c^2 +{p}^2\right) \approx
0,\,\,\,\,\, \,\,\, {p}^2:= {\bf p}^2 -p_0^2.
\label{H1}
 \end{eqnarray}
Indeed, the equations of motion are obtained by taking Poisson
brackets with H:
\begin{eqnarray}
\dot x^\mu=\{x^\mu,\,H\}={1\over\lambda}\,p^\mu,\,\,\, \dot
p^\mu=\{p^\mu,\,H\}=0 \\ \hbox{for}\,\,\,\,\,\,\,
\{x^\mu,\,p_\nu\}= \delta^\mu_\nu.\hskip 3.35truecm \nonumber
\label{Eq1b}
 \end{eqnarray}
Here $\lambda$ is a {\em Lagrange multiplier} (assumed independent
of $x$); it is linked to the choice of a time scale. The
Hamiltonian  constraint gives rise to a singular Lagrangian
through the {\em Legendre transform}, ${\cal L}= {p} \dot {x} - H
$ for ${p}$ determined (as a function of $\dot {x}$) from $\dot
{x} = {{\partial H}\over {\partial  {p}}}$:
\begin{eqnarray}
{\cal L}({x},\dot {x};\lambda)\,\,\,(= {p}\dot {x}
-H)\,\,={\lambda \over 2}\dot {x}^2 - {{m^2c^2}\over {2\lambda}}.
\label{L}
 \end{eqnarray}

{\bf Remark 1} {\em It is important to remember -- especially for
the subsequent extension to a pseudo-Riemannian space-time -- that
the component $p_0$ of the covariant 4-momentum $p_\mu$,
canonically conjugate to the space-time coordinate $x^\mu$, is
{\em minus} the energy, $p_0=-E_p/c$ ($E_p >0$; for a free
particle $E_p/c=\sqrt{m^2 c^2 + \bf p^2}$). The equality $p^0 =
E_p/c$ is not generally covariant, it is accidentally valid for
Cartesian coordinates in flat Minkowski space.}

For $m^2 > 0$ $\lambda$ can be excluded from the condition
\begin{eqnarray}
{{\partial {\cal L}}\over {\partial \lambda}}\!=\!{1\over
2}\!\left(\!\dot {x}^2\! +\!
{{m^2c^2}\over{\lambda^2}}\!\right)\!=\!0 \,\,\Rightarrow\,\,
{\cal L}({x},\dot {x})\!=\! -mc\sqrt{-\dot {x}^2}.
\label{L0}
 \end{eqnarray}

{\bf Remark 2} {\em  For $m=0$ Eq. (\ref{L0}) implies the
constraint $\dot {x}^2\approx 0$ and only the original expression
(\ref{L}) for the Lagrangian remains meaningful.}

For a two particle system we introduce a pair of generalized mass-shell constraints
\begin{eqnarray}
\varphi_a={1\over 2}\left(p_a^2+ m_a^2 + \Phi\right)\approx 0,\,\,\,a=1,2
\label{10phi}
 \end{eqnarray}
satisfying the (strong) compatibility condition
\begin{eqnarray}
\{\varphi_1,\varphi_2\}=\left(p_2{\partial \over {\partial {x_2}}}-
p_1{\partial \over {\partial {x_1}}} \right)\Phi =0.
\label{11phi}
 \end{eqnarray}
Denote by $P$ and $w$ the CM momentum and the total energy:
\begin{eqnarray}
{P}={p}_1+{p}_2,\,\,\,\,\,\,\,\, w^2 = -{P}^2 c^2.
\label{P}
 \end{eqnarray}
We shall exploit the fact that the difference $\varphi_1 - \varphi_2$
of the constraints (\ref{10phi}) is independent of the interaction  $\Phi$ to define the {\em relative
momentum}
\begin{eqnarray}
{p}=\mu_2 {p}_1 - \mu_1 {p}_2, \,\,\,\,\,\,\, \mu_1+\mu_2 = 1
\label{p}
 \end{eqnarray}
determining $\mu_1-\mu_2$ from the strong equation
\begin{eqnarray}
2 P p = 2\varphi_1 - 2\varphi_2 = m_1^2c^2 + {p}_1^2 - m_2^2c^2 -{p}_2^2
\,\,\,\Leftrightarrow
\nonumber\\ \mu_1-\mu_2= {{m_1^2-m_2^2}\over
{w^2}}c^4. \hskip 1.5truecm
\label{pP}
 \end{eqnarray}
(Thus, for unequal masses, the $\mu_a$ depend on the Poincar\'e
invariant total energy square, so that the relation (\ref{p}) is
actually nonlinear.)
The constraint $\varphi_1 - \varphi_2\approx 0$ together with (\ref{pP})
implies the orthogonality of $p$  and $P$ as an universal
{\em kinematical constraint}, readely solved in the
CM frame:
\begin{eqnarray}
{p}{P}\approx 0,\,\,\,\,\left({P}=\left({w/c},\,\bf 0\right)
\,\,\,\Rightarrow\,\,\,{p}=(0,\,\bf p)\right).
\label{pPPp}
 \end{eqnarray}

In order to solve the compatibility equation (\ref{11phi}) we introduce
the projection $x_\perp$ of the relative coordinate  $x_{12}=x_1-x_2$ on  the 3-space
orthogonal to $P$:
\begin{eqnarray}
x_\perp=x_{12} +c^2 {{P x_{12}}\over {w^2}}P.
\label{16x}
 \end{eqnarray}
Its square,
\begin{eqnarray}
R^2 = x_\perp^2=x_{12}^2 +c^2 {{(P x_{12})^2}\over {w^2}}
\label{17R}
 \end{eqnarray}
provides an invariant measure of the distance between the two particles
in the CM frame.

The general Poincar\'e invariant solution of (\ref{11phi}) will be written in the  form
\begin{eqnarray}
\Phi=\Phi\left(R, p^2,p_R;E_w\right),\,\,\,\hbox{where}\,\,\, R p_R = p x_\perp.
\label{10Phi}
 \end{eqnarray}

The effective particle energy $E_w$ is singled out since, as we shall see, $\Phi$
is a quadratic  polynomial in $E_w$ in the cases of interest.
One should also assume that $\Phi \to 0$ for $R \to \infty$, thus making the separation
of the mass terms in (\ref{10phi}) meaningful.

The {\em Hamiltonian constraint} which replaces the equality
$\bf p^2 \approx b^2(w^2)$ (\ref{eq13})
in the presence of interaction is  given by
\begin{eqnarray}
H:= {1\over \Lambda}\left(\mu_2\varphi_1 + \mu_1\varphi_2\right) = \nonumber \\
{1\over {2\Lambda}}\left(p^2 +m_w^2 + \Phi - E_w^2\right) \approx 0.
\label{19H}
 \end{eqnarray}

{\bf Remark 3} {\em The $\mu_a$ defined in (\ref{p}) and
(\ref{pP}) have a simple expression in terms of the  CM energies
$E_a$,
\begin{eqnarray}
-{{c^2}\over w}Pp_1 \approx E_1 = {w\over 2}+
{{m_1^2-m_2^2}\over{2w}}c^4, \nonumber \\
-{{c^2}\over w}Pp_2
\approx E_2 = {w\over 2}- {{m_1^2-m_2^2}\over{2w}}c^4
\label{R23}
 \end{eqnarray}
and approach their non-relativistic values ${m_a\over M}$ for
${w\over {c^2}}\,\,\, \rightarrow \,\,\, M$:
\begin{eqnarray}
\mu_{a}= {E_{a}\over w}= {1\over 2}\pm
{{m_1^2-m_2^2}\over{2 w^2}}c^4\,\,\, \stackrel{{w/{c^2}}\to
M}{-\!\!\!\!-\!\!\!\!-\!\!\!\!-\!\!\!\!-\!\!\!\!-\!\!\!\!-\!\!\!\!\!-\!\!\!\!
\!\!\longrightarrow}\,\,\,{{m_{a}}\over M},\,\,\,a=1,2.
\label{mu12}
 \end{eqnarray}
We note that there are precisely three ways to factorize $b^2$
into $\left({E\over c}-mc\right)\left({E\over c}+mc\right)$ for
$E$ equal to $E_1$, $E_2$ and $E_w$ (\ref{Ew}):
\begin{eqnarray}
b^2(w^2)={{E_a^2}\over {c^2}}-m_a^2c^2={{E_w^2}\over
{c^2}}-m_w^2c^2, \,\,\,\,(a=1,2).
\label{bE}
 \end{eqnarray}
}

They correspond to the ${1\over 2}{4 \choose 2}$ subdivisions of
the four zeros of the numerator of (\ref{eq13}) into two pairs.

This provides a fresh justification for the expressions (\ref{mw}) and (\ref{eq13})
for the relativistic reduced mass and effective particle energy. We shall
present yet another argument in favour of these expressions
starting with a {\em Lorentz covariant} concept of a
{\em relative velocity} (cf. \cite{Fock} Sec.16).

To simplify writing we choose units for which $c=1$. The
4-velocities ${u}_a$ of the two particles and the CM velocity
${U}$ are proportional to the corresponding momenta:
\begin{eqnarray}
{p}_a=m_a {u}_a,\,\,\,a=1,2;\,\,\,\,{P}=w{U},\nonumber\\m_a U
u_a\approx -E_a ,\,\,\,{U}^2=-1.
\label{up}
 \end{eqnarray}
The CM energies $E_1$ and $E_2$ (\ref{R23}) are thus related to the
inner products of the velocities.

We note that the constraints (\ref{up}) are equivalent to the
kinematical constraint (\ref{pPPp}).

The 4-momentum of an {\em effective particle with CM energy}
$\epsilon$,\, ${u}=(u_0,\,{\bf u}), \,\,\, u_0^{CM}= -\epsilon$ is
introduced by defining first the {\em relative 3-velocity} ${\bf
u}$ as follows.

Let  $\Lambda=\Lambda({U})$ be the pure Lorentz transformation
that carries the CM 4-velocity $ {U}$ into its rest frame. The
conditions $\Lambda^\mu{}_\nu U^\nu=\delta^\mu_0$
and positive definiteness determine the (symmetric) Lorentzian matrix
$\Lambda$ uniquely:

\begin{eqnarray}
 \Lambda=\pmatrix{U^0 &-U_j\cr
                 -U^i& \delta^i_j+{{U^iU_j}\over{1+U^0}}\cr
                 }.
\label{LambdaU}
 \end{eqnarray}
We then find that the space parts of $\Lambda u_1$ and $\Lambda
u_2$ are proportional to the same 3-vector ${\bf u}$:
\begin{eqnarray}
{\bf \Lambda u}_1= {m_2 \over w} {\bf u},\,\,\,{\bf \Lambda u}_2=
-{m_1 \over w} {\bf u},
\label{20}
 \end{eqnarray}
which we shall identify with the relative 3-velocity:

\begin{eqnarray}
{\bf u}={{\left(u_2^0+U u_2\right){\bf u}_1 -\left(u_1^0+U
u_1\right){\bf u}_2}\over {1+U^0}}.
\label{pmw}
 \end{eqnarray}

(Note that a similar procedure reproduces the non-relativistic relative
velocity ${\bf v} ={\bf v}_1 -{\bf v}_2$. Indeed, a Galilean
transformation that sends the CM velocity ${\bf
V}={{m_1}\over{M}}\,{\bf v}_1+{{m_2}\over{M}}\,{\bf v}_2$
to zero gives
${\bf v}_1^{CM}={\bf v}_1-{\bf V} ={{m_2}\over M}{\bf v},\,\,\,
{\bf v}_2^{CM}={\bf v}_2-{\bf V} =-{{m_1}\over M}{\bf v}$.)
In particular, in the CM frame we find
\begin{eqnarray}
{\bf P}=m_1{\bf u}_1+m_1{\bf u}_2=0\,\,\,\Rightarrow \nonumber \\
{\bf u}={{E_2}\over {m_2}}{\bf u}_1-{{E_1}\over {m_1}}{\bf
u}_2={w\over{m_2}}{\bf u_1}=-{w\over{m_1}}{\bf u_2}.
\label{22}
 \end{eqnarray}
The time component $u_0$ of the effective particle 4-momentum is
determined by the condition:
\begin{eqnarray}
-U u=-u_0^{CM}= \epsilon.
\label{23}
 \end{eqnarray}

Here $\epsilon$ is given by (\ref{epsilon}); for free particles it
can be interpreted as the common value of the energy component
$-u_{10}$ in the rest frame of $u_2$ and of $-u_{20}$ in the rest
frame of $u_1$ which is a Lorentz invariant; in general, the following
strong equation takes place
\begin{eqnarray}
{u}_1{u}_2\!+\!{{m_1}\over{2m_2}}\left(u_1^2+1\right)
\!+\!{{m_2}\over{2m_1}}\left(u_2^2+1\right)\!=
\\ -\epsilon= {{m_1^2\!+\!m_2^2\!-\!w^2}\over{2m_1m_2}}.\nonumber
\label{eps_}
 \end{eqnarray}

We identify, as usual, the space part of the {\em effective
particle 3-momentum} $\bf p$ in the CM frame with the common value
of ${\bf p}^{CM}_1$ and  $-{\bf p}^{CM}_2$ and {\em define the
proportionality coefficient between $\bf p$ and the relative
3-velocity ${\bf u}$ as the relativistic reduced mass} $m_w$:
\begin{eqnarray}
({\bf p}_{eff}=)\,\,{\bf p}={\bf p}^{CM}_1=-{\bf p}^{CM}_2=
m_1{\bf u}^{CM}_1=m_w{\bf u}
\label{peff}
 \end{eqnarray}
where ${\bf p}^{CM}_1$can also be expressed in terms of the
components of $p_1$ and $p_2$ in an arbitrary frame: $${\bf
p}^{CM}_1={\bf\Lambda p}_1= {{(E_2+p_2^0)\,{\bf
p}_1-(E_1+p_1^0)\,{\bf p}_2}\over{w+P^0}}.$$

(The relation (\ref{peff}) is consistent with the non-relativistic
limit in which ${\bf p}^{CM}=m_1{\bf v}_1=-m_2{\bf v}_2=\mu{\bf
v}$.)

The relativistic reduced mass $m_w$ defined by (\ref{peff}) is
given, as anticipated, by (\ref{mw}). The {\em effective particle
4-momentum} ${p}_{eff}$ is expressed in terms of the relative
momentum ${p}$ (\ref{p}) and the CM 4-velocity ${U}$ by $$
{p}_{eff}= E_w {U} + {p},$$ \vskip -.6truecm
$$\hbox{i.e.}\,\,\,\,\,\,\,{p}^{CM}_{eff}=(-E_w,\,{\bf p}),\,\,\,
{p}^{CM}=(0,\,{\bf p}).\hskip 1.6truecm$$

For positive mass particles it is convenient to write the
Hamiltonian constraint (\ref{19H}) in terms of a dimensionless
4-momentum, corresponding to setting $\Lambda = \lambda m_w^2
c^4\,\left(=\lambda m_w^2\right)$:
\begin{eqnarray}
\hat H= {1\over{2\lambda}}({\bf u}^2 +\hat \Phi  + 1
-\epsilon^2)\approx 0,
\label{H0u}
 \end{eqnarray}
\begin{eqnarray} \hat\Phi = {\Phi \over {m_w^2 c^4}}=\hat\Phi\left(r, {\bf
u}^2, u_r;\epsilon\right)
\label{H0u2}
 \end{eqnarray}
where the radial normalized momentum $u_r$ is dimensionless while r has dimension
of an action: $r= R m_wc$, $p_R=m_wcu_r$.

The interaction function $\hat\Phi$ in (\ref{H0u})
is chosen to reproduce the known interaction
of a test particle  in an external field. For the combined
electromagnetic and gravitational interaction of
two particles of charges $e_1$ and $e_2$ we shall write:
\begin{eqnarray}
H={1\over {2\lambda}}\left(\!1\!+\!\left(1\!-\!2{{\alpha_{{}_G}}\over r}\right)u_r^2 \!+\!
{{J^2}\over {r^2}}\!-\!
{{\left(\epsilon\!-\!{{e_1 e_2}\over {c r}}\right)^2}\over {1\!-\!2{{\alpha_{{}_G}}\over r}}} \right)
\approx 0 \nonumber \\
\hbox{for}\,\,\, r> 2\alpha_{{}_G}
\label{34H}
 \end{eqnarray}
where $J^2$ is the square of the total angular momentum while $\alpha_{{}_G}$
(denoted in \cite{BD99} by $\alpha$)
is the gravitational coupling measured in units of action:
\begin{eqnarray}
{\bf J}={\bf r}\times {\bf u},\,\,\, {\bf r}= m_wc{\bf x}_\perp^{CM},\,\,\,
r u_r={\bf r}{\bf u},\nonumber \\r^2={\bf r}^2,\,\,\, J^2={\bf J}^2;
\,\,\, \alpha_{{}_G}=G{{m_1 m_2}\over c}.
\label{35J}
 \end{eqnarray}
This Hamiltonian constraint corresponds to the dimensionless interaction function
\begin{eqnarray}
\hat\Phi\!=
\!-\!2{{\alpha_{{}_G}}\over{r\!-\!2\alpha_{{}_G}}}\epsilon^2
\!-\!2{{\hbar\alpha}\over{r\!-\!2\alpha_{{}_G}}}\epsilon
\!-\!{{\hbar^2\alpha^2}\over{r\left(r\!-\!2 \alpha_{{}_G}\right)}}
\!-\!2{{\alpha_{{}_G}}\over r}u_r^2, \nonumber \\
\alpha = - {{e_1e_2}\over{c\hbar}} \hskip 3.5truecm
\label{36Phi}
 \end{eqnarray}
($\alpha$ is  positive for oppositely charged particles,
studied in Sec.3 below; it coincides  with the
fine structure constant , $\alpha^{-1}=137.036$,
for $e_1=-e_2$ equal to the electron charge).
It is remarkable that the Poincar\'e
invariant constraints in flat phase space admit an interpretation
in terms of an effective particle moving
along geodesics in a Schwarzschild space-time.

An advantage of our choice of $\Lambda$ and of the variables ${\bf r}$, ${\bf u}$ and
$\epsilon$ (instead of ${\bf R}= {\bf x}_\perp^{CM}$, ${\bf p}^{CM}$, and $E_w$)
yielding the dimensionless Hamiltonian constraint (\ref{34H}) is the quadratic dependence of $H$
in the (single !) energy parameter $\epsilon$ (instead of the two $w$-dependent quantities $E_w$
and $m_w$ in (\ref{19H}).
This allows to write down a Lagrangian for the (interacting) two-particle system
using the standard Legendre transform
\begin{eqnarray}
{\cal L}({\bf r}, \dot t, \dot r, \dot\varphi,;\lambda)=
{\bf u}\dot{\bf r}-\epsilon\dot t -H, \,\,\,{\bf u}\dot{\bf r}= u_r\dot r
+J \dot\varphi, \nonumber \\
\epsilon\!=\!\lambda\left(1\!-\!2{{\alpha_{{}_G}}\over{r}}\right)\dot t\!-\!{{\hbar\alpha}\over r},
\,\,\,u_r\!=\!{{\lambda \dot r}\over{1\!-\!2{{\alpha_{{}_G}}\over r}}},\,\,\,J\!=\!\lambda r^2\dot\varphi
\label{37L}
 \end{eqnarray}
yielding
$${\cal L}=
{\lambda\over 2}
\left(
{{\dot r^2}\over{1-2{{\alpha_{{}_G}}\over r}}}
+ r^2\dot\varphi^2-
\left(1-2{{}\alpha_{{}_G}\over r}\right)\dot t^2
\right) +
\dot t {{\hbar \alpha}\over r}- {1\over {2\lambda}},$$
or, varying in $\lambda$ and excluding it from the resulting constraint,
\begin{eqnarray}
{\cal L}= \dot t {{\hbar\alpha}\over r}- \lambda ^{-1}=\nonumber \\
\dot t{{\hbar\alpha}\over r}-
\left(
\left(1-2{ {\alpha_{{}_G}}\over r}\right)\dot t^2
-{{\dot r^2}\over{1-2{{\alpha_{{}_G}}\over r} }} -r^2 \dot\varphi^2
\right)^2.
\label{38L}
 \end{eqnarray}

Here we have used angular momentum conservation which implies that tha effective particle moves in a plane
orthogonal to ${\bf J}$:
\begin{eqnarray}
{\bf r}= r(\cos \varphi, \sin\varphi, 0)\,\,\,\hbox{for}\,\,\, {\bf J}=(0,0,J).
\label{39r}
 \end{eqnarray}

We shall study the case of electromagnetic and gravitational
interaction (corresponding to $\alpha_{{}_G}=0$ and to
$\hbar\alpha = 0$, respectively) in Sec.3 and 4 below solving in
each case the resulting equations of motion. The result for the
first relativistic ("post-Newtonian") approximation agrees with
(more complicated) traditional calculations. Higher order
corrections require taking into account retardation effects which
can be also done within  the present 1-body approach, as discussed
in Sec.5. The Wheeler-Feynman non-local action for electrically
charged particles \cite{WF49} (see also \cite{K72}) seems to
provide a systematic treatment of retardation effects (to all
orders) but its inclusion in the present framework is not obvious.

\section{The bound-state problem for two oppositely charged relativistic particles}

We start with the Hamiltonian constraint (\ref{34H})
in the absence of gravitational forces (i.e. with $G=0=\alpha_{{}_G}$):
\begin{eqnarray}
2\lambda H = u_r^2+{{J^2}\over{r^2}}+1-\left(\epsilon +{{e^2}\over
r}\right)^2\approx 0, \nonumber \\
e^2\equiv\hbar\alpha=-{{e_1e_2}\over c}.
\label{HJr_em}
 \end{eqnarray}

The canonical Poisson bracket relations
$\{x_a^\mu,p_{b\nu}\}=\delta_{ab}\delta_{\mu\nu}$
imply the following non-zero brackets for the radial and angular
variables relevant for the planar motion:
\begin{eqnarray}
\{r,\,u_r\}=1=\{\phi,\,J\}.
\label{Pbr_em}
 \end{eqnarray}
The equations of motion derived from (\ref{HJr_em}),
(\ref{Pbr_em}) read:
\begin{eqnarray}
\dot r={{\partial H}\over {\partial u_r}}= {u_r\over
\lambda},\,\,\,\dot\phi={{\partial H}\over {\partial J}}= {J\over
{\lambda}r^2},\,\,\,\dot J={{\partial H}\over {\partial\phi}}= 0.
\label{EqM_em}
 \end{eqnarray}
The equation for the effective particle trajectory obtained by
dividing $\dot r$ by $\dot\phi$ is independent of $\lambda$:
\begin{eqnarray}
-{{d}\over{d\phi}}\left({J\over r}\right) = u_r = \sqrt{
2\epsilon\alpha_{{}_J}{J\over r}
-\left(1-\alpha_{{}_J}^2\right)\left({J\over r}\right)^2-\beta }
\label{EqTr_em}
 \end{eqnarray}
where all variables -- starting with ${J\over r}$ -- are
dimensionless, $\alpha_{{}_J}$ plays the role of a "classical fine
structure constant":
\begin{eqnarray}
\alpha_{{}_J}={{e^2}\over J}\,\left(={{|e_1 e_2|}\over {c
J}}={\hbar\over J}\alpha\right).
\label{alphaJ}
 \end{eqnarray}

 The bounded motion corresponds to the case when the
expression under the square root has two positive zeros in
${J\over r}$. This implies
\begin{eqnarray}
0<\beta\equiv 1-\epsilon^2\leq \alpha_{{}_J}^2,
\label{cond1}
 \end{eqnarray}
\vskip -.5truecm $$i.e.,\,\,\,\,\,\,0< -b^2(w^2)\leq
\alpha_{{}_J}^2 m_w^2 .\hskip 1.4truecm $$

In fact, introducing the dimensionless inverse radius variable
\begin{eqnarray}
y={ {1-\alpha_{{}_J}^2} \over {\epsilon\,\alpha_{{}_J}}}\,{J\over
r},
\label{41}
 \end{eqnarray}
we find
\begin{eqnarray}
\left({{d y}\over {d \phi}}\right)^2 =
\left(1-\alpha_{{}_J}^2\right) \left(-{{1-\alpha_{{}_J}^2}\over
{\alpha_{{}_J}^2}}{\beta\over {1-\beta}} +2 y -y^2\right).
\label{dy}
 \end{eqnarray}

 The discriminant of the quadratic expression in $y$ in the
right hand side is positive whenever Eq. (\ref{cond1}) takes
place. (The positivity requirement for the two real roots in
${J\over r}$ implies that $1-\epsilon^2$ and $1-\alpha_{{}_J}^2$
are both positive.) We can then rewrite Eq. (\ref{EqTr_em}) in the
form \begin{eqnarray} {{d
y}\over{\sqrt{(y_1-y)(y-y_2)}}}=-\sqrt{1-\alpha_{{}_J}^2}\, d \phi
\label{313} \end{eqnarray} $$\hbox{with}\,\,\,\,\,\,\,\,\,\, y_{1,2}=1\pm
e(\beta,J)\hskip 2.truecm $$ where $e(\beta,J)$ plays the role of
eccentricity: \begin{eqnarray}
e^2(\beta,J)=1-{{1-\alpha_{{}_J}^2}\over
{\alpha_{{}_J}^2}}{\beta\over {1-\beta}} ,\,\,\,\,\,\,0\leq
e(\epsilon,J)<1, \label{ecc_em} \end{eqnarray} the last inequality
being valid in the domain (\ref{cond1}), i.e. for  $
1<\alpha_{{}_J}^2+\epsilon^2$,\,\, $0<\alpha_{{}_J}<1$,\,\,
$0\leq\epsilon<1$ in the $(\alpha_{{}_J},\epsilon)$ plane.

Integrating Eq. (\ref{313}) with initial condition $y(\phi=0)=y_1$
(i.e., the orbit passes through the perihelion for $\phi=0$) we
obtain \begin{eqnarray}
y=1+e(\beta,J)cos\left(\sqrt{1-\alpha_{{}_J}^2}\,\,\phi\right).
\label{315} \end{eqnarray}

To compare this result with the familiar non-relativistic elliptic
orbit we first observe that for $w$ given by (\ref{w}) $\epsilon$
(\ref{epsilon}) is expressed in terms of the dimensionless measure
$\varepsilon$ of the binding energy and the ratio $\nu$ between
the reduced and the total mass,
\begin{eqnarray}
\varepsilon={ E \over {\mu c^2}}\,\,\,\left(\mu={{m_1m_2}\over M
}\right), \,\,\,\,\,\,\nu={\mu\over
M}\,\,\,\,\,\,(|\varepsilon|\ll 1),
\label{316}
 \end{eqnarray}
as follows
\begin{eqnarray}
\epsilon=1+\varepsilon+{\nu\over 2}\varepsilon^2
\label{317}
 \end{eqnarray}
so that
\begin{eqnarray}
\beta = 1-\epsilon^2=
-\varepsilon\,\Bigl(2+(\nu+1)\varepsilon\Bigr)+O(\varepsilon^3).
\label{betaepsilon}
 \end{eqnarray}
In the non-relativistic limit Eqs. (\ref{41}) and (\ref{315})
yield an elliptic trajectory
\begin{eqnarray}
y_{{}_{NR}} = { J \over { \alpha_{{}_J} r}} = 1 +
\sqrt{1+{{2\varepsilon_{{}_{NR}}}\over{\alpha_{{}_J}^2}}}\cos\phi.
\label{yNR}
 \end{eqnarray}

The relativistic orbit (\ref{315}), on the other hand, is not
closed (except for $e(\beta,J)=0$). The perihelion shift
$\delta\phi$ is given by
\begin{eqnarray}
\delta\phi\!=\!2\pi\!\left(\!\left(\!1-\!\alpha_{{}_J}^2\!\right)^{-{1/2}}
\!-\!1\!\right)\!\!=\!\pi\alpha_{{}_J}^2\!+\!O(\alpha_{{}_J}^4)
\,\,\,\hbox{for}\,\,\,\alpha_{{}_J}^2\ll 1.
\label{p_shift_em}
 \end{eqnarray}

For a circular orbit we have
$$e^2(\beta,J)=0\,\Rightarrow\,\beta=\alpha_{{}_J}^2,\,\,\,\,
i.e.,\,\,\,-\varepsilon=f(\alpha_{{}_J}^2,\nu)$$
$$f(\alpha_{{}_J}^2,\nu)\!=\!{1\over
\nu}\!\left(\!1\!-\!\sqrt{1\!-\!2\nu\left(1\!-\!\sqrt{1\!-\!\alpha_{{}_J}^2}\right)}
\right)=$$ \vskip -.3truecm $${1\over
2}\alpha_{{}_J}^2\!+\!{{\nu\!+\!1}\over 8}\alpha_{{}_J}^4
\!+\!O(\alpha_{{}_J}^6).$$

In general, for a bounded motion, we have the inequality
\begin{eqnarray}
0<-\varepsilon\leq f(\alpha_{{}_J}^2,\nu).
\label{varepsilon}
 \end{eqnarray}

 These results -- and their derivation --
should be compared with the conventional approach that starts with
the approximate Hamiltonian (see \cite{LL}, Sec.65, Problem 2):
\begin{eqnarray} H={{{\bf p}^2}\over{2\mu}}\left(1-{{1-3\nu}\over{4}}{{{\bf
p}^2}\over{\mu^2c^2}}\right) - {{e^2}\over R}\left(1 + \nu\,
{{{\bf p}^2 + p_R^2}\over{2\mu^2 c^2}}\right) \label{320}
\end{eqnarray} \vskip -.7truecm $$\hbox{for}\,\,\,\,\,\,\,\, {\bf
p}^2=p_R^2+{{J^2}\over{R^2}}\,\,\,\left(R=|{\bf
x}_{12}^{CM}|\right).\hskip 1.5truecm $$

A computation which involves a redefinition of $R$ according to
the substitution $R^2 \rightarrow
{R^\prime}^2=R\left(R-{{e^2}\over{Mc^2}}\right)$, and is certainly
less transparent then the above, yields a trajectory whose
parameters agree with (\ref{315}) up to (including) order
$\alpha_{{}_J}^2$ (Velin G. Ivanov, Diploma Work, Sofia, 2000).

\section{Gravitational 2-body problem}

The {\em Hamiltonian constraint for the
gravitational interaction} of two (point) particles of {\em
arbitrary masses} $m_1, m_2$, obtained from (\ref{34H}) for
$e_1e_2=0$, can be interpreted as the condition that the effective
particle 4-velocity ${u}=(\epsilon,{\bf u})$ has unit mass in a
Schwarzschild metric whose "radius" $2\alpha_{{}_G}=m_w c\,R_w$
(of dimension of an action) is determined by the two masses:
\begin{eqnarray}
H={1\over{2\lambda}}\left( 1+g^{00} \epsilon^2 +g^{ij} u_i
u_j\right)\approx 0.
\label{H_gr}
 \end{eqnarray}
(A constraint of this type has been first written in \cite{MNT81}
where, however, a more complicated metric was introduced, computed
in a quantum field theoretic framework. As observed in \cite{BD99}
the classical Schwarzschild metric gives a better approximation --
in accord with our general prescription of Sec.2.) Here the metric
is expressed in terms of the radial variable $r$ by
\begin{eqnarray}
g_{00}= 2{ {\alpha_{{}_G}} \over r}-1 =
{1\over{g^{00}}}(<0),\nonumber
\\ g^{ij} u_iu_j=\left(1-2{{\alpha_{{}_G}}\over r}\right)u_r^2
+{{J^2}\over{r^2}}
\label{gSch}
 \end{eqnarray}
where $r$, $u_r$,  $J$ and $\alpha_{{}_G}$ are given by (\ref{35J}).

Proceeding to the Hamiltonian equations of motion we introduce the
($J$ dependent) dimensionless coupling parameter
\begin{eqnarray}
\rho\,(=\rho_{{}_J})= {{2\alpha_{{}_G}}\over J}= {{R_w m_w c}\over
J }={{2 m_1 m_2 G}\over {c J}}
\label{rho}
 \end{eqnarray}
($R_w=2{{Gw}\over{c^4}}$ being the energy dependent "Schwarzschild
radius"). As we shall see shortly, $\rho^2 <{1\over 3}$ for the
bounded motion; by contrast, the counterpart $\alpha_{{}_G}/\hbar$
of the electromagnetic fine structure constant is rather big: for
$m_1=m_2=M_\odot$ (the solar mass) it is of the order of
$10^{76}$.(The parameter $\rho$ coincides with $2/j$ of
\cite{BD99}.)

The Hamiltonian constraint can be written in terms of $\rho$ and
${J\over r}$ as
\begin{eqnarray}
H={1\over{2\lambda}}\left(1+\left(1-\rho\,{{J}\over r}\right)u_r^2
+{{J^2}\over {r^2}}- {{\epsilon^2}\over{1-\rho\,{{J}\over
r}}}\right)\approx 0.
\label{H_rJ_gr}
 \end{eqnarray}

 The Poisson brackets (\ref{Pbr_em}) remain unchanged and we
deduce as before the equations of motion
\begin{eqnarray}
\dot r={{\partial H}\over{\partial
u_r}}=\left(1\!-\!\rho\,{{J}\over r}\right)
{{u_r}\over\lambda},\,\,\,\,\,\,\dot \phi={{\partial
H}\over{\partial J}}= {J\over{\lambda r^2}}.
\label{EqM_gr}
 \end{eqnarray}

Introducing again a dimensionless variable proportional to the
inverse radius (cf. (\ref{41})), \begin{eqnarray} y={J\over r}
\label{410} \end{eqnarray} we obtain the following ($\lambda$-independent)
differential equation for the effective particle trajectory:
\begin{eqnarray} -{{d y}\over{d\phi}}=(1-\rho y)u_r=\left(\rho
y^3-y^2+\rho y -\beta \right)^{1/2}, \label{trajectory_gr}
\end{eqnarray} \vskip -.7truecm $$\beta=1-\epsilon^2.$$

The energy independent coefficient $\rho=\rho_{{}_J}$ will play
the role of dimensionless expansion parameter (replacing the
commonly used ${1\over c}$).

Eq. (\ref{trajectory_gr}) can be solved in terms of Jacobi
elliptic functions (cf. \cite{Singe} Sec.VII.8). To begin with, we
assume that all three zeros $y_0, y_1, y_2$ of the cubic
polynomial under the square root are positive reals
\begin{eqnarray}
P_3(y)\!:=\!\rho y^3\!-\!y^2\!+\!\rho y
\!-\!\beta\!=\!\rho(y\!-\!y_0)(y\!-\!y_1)(y\!-\!y_2),\\ 0<y_2\leq
y_1<y_0.\nonumber
\label{roots}
 \end{eqnarray}

The finite (bound state) motion belongs to the range $y_2\leq
y\leq y_1$ for which $P_3(y)$ is non-negative. (The infinite interval
$y>y_0$, in which $P_3(y)>0$ as well, corresponds to falling on a
centre.) The necessary and sufficient conditions for $\rho$ and
$\beta$ for which all zeros of $P_3$ are positive are the
positivity of $\beta$,
\begin{eqnarray}
 0<\beta\,(=1-\epsilon^2)<1,\,\,\, \hbox{i.e.,}\,\,\,
(m_1^2+m_2^2<)\,{{w^2}\over{c^4}}<M^2
\label{413}
 \end{eqnarray}
and the non-negativity of the discriminant:
\begin{eqnarray}
(0<)\,\rho^2\!\leq\!{1\over 3},\,\,\, 27\left(\!\beta \rho^2
\!+\!{1\over 3}\!\left(\!{2\over
9}\!-\!\rho^2\!\right)\!\right)^2\!\leq\!4\left(\!{1\over
3}\!-\!\rho^2\!\right)^3\!.
\label{414}
 \end{eqnarray}
We begin our discussion with the case of vanishing discriminant
that gives rise to a {\em circular orbit} with
\begin{eqnarray}
y_1=y_2=:y_c={{1-\sqrt{1-3\rho^2}}\over{3\rho}}=\nonumber\\
{\rho\over 2}+{3\over 8}\rho^3+{9\over
{16}}\rho^5+O(\rho^7),\nonumber\\ y_0={1\over \rho}-2 y_c
\label{roots_c}
 \end{eqnarray}
obtained by solving the quadratic equation ${{d P_3}\over{d y}}=0$
for
\begin{eqnarray}
\beta\!=\!(\rho y_c^2\!-\!y_c\!+\!\rho)y_c\!=\!
{2\over{27\rho^2}}\!\left(\!\left(\!1\!-\!3\rho^2\!\right)^{3/2}
\!-\!1+\!{9\over 2}\rho^2\!\right)\!=\! \nonumber\\{{\rho^2}\over
4}+{{\rho^4}\over 8}+{9\over{64}}\rho^6+O(\rho^8).
\label{416}
 \end{eqnarray}

A measure of the frequency on a circular orbit is
\begin{eqnarray}
\omega={{d\phi}\over{d t}}=\dot\phi/\dot t={J\over{\epsilon
r^2}}\left(1-\rho\,{{J}\over r}\right)={{y_c^2}\over{\epsilon
J}}\left(1-\rho\,y_c\right)
\label{omega}
 \end{eqnarray}
where we have used (\ref{H_rJ_gr}), (\ref{EqM_gr}) and (\ref{410}).

The {\em last} (innermost) {\em stable circular orbit}, LSO (whose
significance stems from the fact that gravitational radiation
damping tends to circularize binary orbits \cite{CE77} -- see also
\cite{BD99} for a discussion and references) corresponds to the
values
\begin{eqnarray}
\rho=\rho^*={1\over\sqrt{3}} = y_c^* = y_0^*,\,\, \beta^*={1\over
9},\,\,\epsilon^*={\sqrt{8}\over 3}, \hskip .6truecm \nonumber \\
{{w^*}\over{c^2}}\!=\!M
\sqrt{1\!-\!2\nu\!\left(\!1\!-\!{\sqrt{8}\over 3}\right)},
\,\,\,m_w^*\!=\!{\mu \over
\sqrt{1\!-\!2\nu\!\left(\!1\!-\!{\sqrt{8}\over 3}\right)} }
\nonumber\\
\left(\alpha_{{}_G} \omega^*\right)^{2/3}\!\!\!=\!{1\over 6},
\hskip 2.2truecm
\label{LSO}
 \end{eqnarray}
for which both sides of the last inequality (\ref{414}) vanish and
all three zeros of $P_3$ coincide. These values correspond to a
limit point of local minima of what is called the "effective
potential" $V(r,J)$ that enters the expression for $\epsilon^2$
obtain from (\ref{H_rJ_gr}):
\begin{eqnarray}
\epsilon^2\left(\!1\!-\!\left({{d r}\over{d
t}}\right)^{\!2}\!\right)=V(r,J):= \left(\!1\!-\!\rho\,{{J}\over
r}\!\right) \left(\!1\!+\!{{J^2}\over{r^2}}\!\right)=
\label{419}
 \end{eqnarray}
\vskip -.6truecm $$\epsilon^2 -P_3(y).$$

We have
\begin{eqnarray}
{d\over{d r}}V(r,J)\!=\!{J\over{r^2}}{{dP_3(y)}\over{d
y}}\put(2,-10){\line(0,1){25}}_{\,\,\,y={J\over
r}}\!\!=\!{{y^2}\over J}\left(3\rho
y^2\!-\!2y\!+\!\rho\right)\!=\!0.
\label{420}
 \end{eqnarray}
It is the smaller zero of the second factor, $y_c$
(\ref{roots_c}), that corresponds to a minimum of $V$. The
distance $r_c={J\over{y_c}}$ decreases when $\rho=\rho_{{}_J}$
increases and attains its minimal value for the maximal possible
value $\rho^*$ (\ref{LSO}) of $\rho$.

The {\em dimensionless binding energy}
$\varepsilon={1\over\mu}\left({w\over{c^2}}-M\right)= -(m_w-\mu)/
\nu m_w$ of the LSO, evaluated from (\ref{317}) and (\ref{LSO})
is:
\begin{eqnarray}
\varepsilon^*= -{1\over
\nu}\left(1-\sqrt{1-2\nu(1-\epsilon^*)}\right)= \nonumber \\
-(1\!-\!\epsilon^*)\left(1\!+\!\nu{{1\!-\!\epsilon^*}\over 2}\!+\!
\nu^2{{(1\!-\!\epsilon^*)^2}\over 2}\right)
\!+\!O\left(\nu^3(\epsilon^*)^3\right),
\label{binE}\\
1-\epsilon^*=1-{\sqrt{8}\over 3}\sim {2\over {35}}.\nonumber
 \end{eqnarray}

The increase of $|\varepsilon^*|(=-\varepsilon^*)$ (compared
to its Schwarzschild value $|\varepsilon^*|=1-\epsilon^*$ ) by the
factor $\left(1+\nu {{1\!-\!\epsilon^*}\over 2}+\dots\right)$ is
coupled to a similar increase of the relativistic reduced mass
\begin{eqnarray}
m_w={{m_1m_2}\over{M(1+\nu \varepsilon^*)}}=
{\mu\over{1-\nu|\varepsilon^*| }}
\label{mm}
 \end{eqnarray}
\vskip -.5truecm $$\hbox{for}\,\,\,{{w^*}\over{c^2}}=
M(1\!+\!\nu\varepsilon^*).$$

The expansion parameter $x=\left(\alpha_{{}_G}\omega
\right)^{2/3}$ of \cite{DJS00} takes for the LSO its Schwarzschild
value $x^*={1\over 6}$ -- see (\ref{LSO}). Returning to the "true
radius" $R={r\over{m_wc}}$ (measured in units of length) of LSO we
find, in accord with \cite{DJS00}, that it is smaller than its
Schwarzschild value $R_{{}_S}^*$
\begin{eqnarray}
R^*\!=\!R^*_{{}_S} {w^*\over{Mc^2}}\!=\!R^*_{{}_S}(1\!+\!\nu
\varepsilon^*)\!\sim\!
R^*_S\!\left(\!1\!-\!\nu\left(1\!-\!{\sqrt{8}\over 3
}\right)\!\right)\!.\!
\label{422}
 \end{eqnarray}

Accordingly the angular frequency $\omega^*$ for the relativistic
two-body LSO is bigger than its Schwarzschild value $\omega_S^*$:
\begin{eqnarray}
\omega^*=\omega_S^* {{Mc^2}\over {w^*}}= {{\omega_S^*}\over{1+\nu
\varepsilon^*}}.
\label{omega*}
 \end{eqnarray}

We now proceed to the general case in which the relation
(\ref{416}) between $\beta$ and $\rho^2$ becomes an inequality
\begin{eqnarray}
0<\beta\leq{2\over{27\rho^2}}\left(\left(1-3\rho^2\right)^{3/2}-1+{9\over
2 }\rho^2\right)\leq \nonumber \\{1\over 4}\rho^2 + {1\over
8}\rho^4+{9\over{64}}\rho^6+{{45}\over{64}}\rho^8\leq{1\over
3}\rho^2 \hskip .35truecm
\label{423}
 \end{eqnarray}
\vskip -.5truecm $$\hbox{for}\,\,\,\,\,\,0<\rho^2\leq{1\over 3}.$$
(The coefficient to $\rho^8$ is chosen in a such a way that for
$\rho^2={1\over 3}$ inequalities (\ref{423}) become equalities.)
We shall compute the zeros of $P_3$ as expansions in $\rho^2$
taking into account the fact that $\beta$ is, according to
(\ref{423}), (at most) of order $\rho^2$.

For the largest root one finds the following expansion
\begin{eqnarray}
y_0={1\over \rho}-\rho\left(1+\rho^2-\beta
+2\rho^4-3\rho^2\beta\right) + O(\rho^7).
\label{424}
 \end{eqnarray}

The two smaller roots are then computed from the relations
\begin{eqnarray}
y_1+y_2={1\over\rho}-y_0=\hskip 2.7truecm\nonumber \\
\rho\left(1+\rho^2-\beta
+2\rho^4-3\rho^2\beta\right)+O(\rho^7),\nonumber \\ \rho\, y_0 y_1
y_2=\beta.\hskip 3.7truecm
\label{425}
 \end{eqnarray}

The result is
\begin{eqnarray}
y_{1,2}= {\rho\over 2}
\left(1+\rho^2-\beta+2\rho^4-3\rho^2\beta\pm\lambda\right)+O(\rho^7),\nonumber\\
\lambda\geq 0,
\label{426}
 \end{eqnarray}
\begin{eqnarray}
\rho^2\lambda^2\!=\hskip 6.3truecm \nonumber \\
\rho^2\!-\!4\beta\!+\!2\rho^4\!-\!6\rho^2\beta\!
+\!5\rho^6\!-\!16\rho^4\beta\!+\!5\rho^2\beta^2\!+\!O(\rho^8).
\label{427}
 \end{eqnarray}
The solution of Eq. (\ref{trajectory_gr}) satisfying $y(0)=y_1$ is
expressed in terms of the elliptic sine function (see
\cite{HTF}):
\begin{eqnarray}
{{y(\phi)-y_2}\over{y_1-y_2}}=sn^2\left(
K-\sqrt{\rho(y_0-y_2)}\,{\phi\over 2}, k\right).
\label{sol}
 \end{eqnarray}
The {\em module square}, $k^2$, of the elliptic functions is
expressed as a ratio of differences of roots of $P_3$:
\begin{eqnarray}
k^2\!=\!{{y_1\!-\!y_2}\over{y_0\!-\!y_2}}\!=\!{{\lambda\rho^2}\over{1\!-\!{3\over
2}\rho^2(1\!+\!\rho^2\!-\!\beta)\!+\!{1\over 2}\rho^2\lambda}}
\!+\!O(\rho^8)\!=\!
\label{k}
 \end{eqnarray}
\vskip -.3truecm $$\lambda\rho^2\left(1+{3\over 2}\rho^2 +
4\rho^4-{5\over 2}\rho^2\beta\right)-$$ \vskip -.5truecm
$${\rho^2\over 2}\left(\rho^2-4\beta
+5\rho^4-18\rho^2\beta\right)+O(\rho^8);$$

$4K(k^2)$ is the {\em real period} of $sn(x,K)$ and $cn(x,k)$:
$sn(K,k)=1, cn(K,k)=0$,\,

\begin{eqnarray}
K\!=\!\int_0^1\!\!{ {d x}\over {
\sqrt{(1-x^2)(1-k^2x^2)}}}\!=\!{\pi\over 2}F\left({1\over
2},{1\over 2},;1;k^2\right).
\label{K}
 \end{eqnarray}

We now proceed to computing the {\em perihelion shift}
$\delta\phi$ of the effective particle. The change $\Delta\phi$ of
$\phi$ for a full turn and the shift $\delta\phi$ are given by
\begin{eqnarray}
\Delta\phi=2\int_{y_2}^{y_1}{{d
y}\over{\sqrt{P_3(y)}}}={{4K}\over{\sqrt{\rho(y_0-y_2)}}}=2\pi+\delta\phi.
\label{Dshift}
 \end{eqnarray}

This is an exact formula. For small  $\rho$ the elliptic module
$k$ is also small (according to (\ref{k})) and we can approximate
$4K$, using (\ref{K}), by
\begin{eqnarray}
4K = 2\pi \left(1+ {{k^2}\over 4}+ {9\over{64}}k^4+ O(k^6)\right)
= \nonumber\\ 2\pi \left(1+{\lambda\over 4}\rho^2 \left(1+{3\over
2}\rho^2\right)+ {{\rho^2}\over
{64}}\left(\rho^2-4\beta\right)+O(\rho^6)\right).
\label{432}
 \end{eqnarray}
Combining this with the expansion
\begin{eqnarray}
\left(\rho(y_0-y_2)\right)^{-1/2}=\hskip 4.6truecm \nonumber \\
\left(1\!-\!{3\over
2}\rho^2(1\!+\!\rho^2-\beta)\!+\!{{\rho^2}\over
2}\lambda\right)^{-1/2} \!+\!O(\rho^6)=\hskip 1truecm \nonumber
\\ 1\!+\!{3\over 4}\rho^2\left(\!1\!+\!{3\over
4}(3\rho^2\!-\!2\beta)\!\right)\!-\!{{\lambda\rho^2}\over
4}\left(1\!+\!{9\over 4}\rho^2\right)\!+\!O(\rho^6)
\label{433}
 \end{eqnarray}
we end up with an expression for $\delta \phi$ in which the odd
powers of $\lambda$ cancel out:
\begin{eqnarray}
\delta\phi={{4K}\over{\sqrt{\rho(y_0-y_2)}}}-2\pi=\hskip 1.6truecm
\nonumber
\\ {{3\pi}\over
2}\rho^2\left(1+{5\over{16}}(7\rho^2-4\beta)\right)+O(\rho^6).
\label{dshift}
 \end{eqnarray}

To compare with earlier calculations \cite{DS88}, \cite{OK89},
\cite{SW93} one again uses the expansion (\ref{betaepsilon}) of
$\beta$ in terms of the dimensionless binding energy (\ref{316}).
The result clearly agrees with the first post-Newtonian
approximation. The missing fourth order (in $\rho$) term
${{3\pi}\over 4}\nu \rho^2\left(\beta -{5\over 4}\rho^2\right)$
can be shown to correspond to retardation effects.

\section{Concluding remarks}

Formulae for particle trajectories in a relativistic 2-body system
(including recoil effects) have been derived with the same ease as
for a test particle problem in a Coulomb or Schwarzschild
potential. The expression $\epsilon = 1 +\varepsilon+{\nu\over
2}\varepsilon^2$ (\ref{317}) for the CM energy per unit mass
$\epsilon={{E_w}\over{m_wc^2}}$ in terms of the dimensionless
measure $\varepsilon={E\over {\mu c^2}}$ of the binding energy
($\varepsilon<0$ for finite motion) has been deduced as a
straightforward consequence of the relation (\ref{w}) between the
CM energy, the total mass $M=m_1+m_2$ and $E$:
\begin{eqnarray}
{w \over{Mc^2}} = 1+\nu \varepsilon,\,\,\,\,\,\, \nu={\mu\over M}=
{{m_1m_2}\over{(m_1+m_2)^2}}.
\label{51}
 \end{eqnarray}
(This simple and natural derivation should be compared with the
rather involved argument of Sec.4 of \cite{BD99} yielding the same
result. It thus provides {\em a posteriori\,} justification of our
definition of a relativistic effective particle -- on top of the
arguments presented in Sec.2.)

A systematic way to compute higher order corrections has been
worked out in the quantum case \cite{T73,RTA75}, and can, in
principle, be applied to the classical limit as well. It is,
however, desirable to have a consistent classical algorithm for
calculating retardation effects.

It is known (and follows by comparing results of the previous
sections with earlier calculations) that these first contribute to
order $\varepsilon^2$ (i.e. to $\alpha_{{{}_J}}^4$ in the
electromagnetic case or to $\rho^4\, (\sim {1\over {c^4}})$ in the
gravitational case). It turns out that it is quite feasible to
include such corrections within the effective 1-body approach
developed here. We shall indicate how to do this by modifying the
effective particle Lagrangian.

It follows from Eq. (26.23) of \cite{Fock} (or from Eq. (65.7) of
\cite{LL}) that the retardation effect in electrodynamics (to
order ${1\over{c^2}}$) is accounted for by multiplying the
interaction term (i.e.,  ${{e^2}\over r}$) by the velocity
dependent expression
\begin{eqnarray}
k_{{}_{EM}}=1-{1\over 2}\left({\bf v}_1{\bf v}_2+{1\over r^2}({\bf
v}_1{\bf r})({\bf v}_2{\bf r})\right)
\label{kEM}
 \end{eqnarray}
(where we are using our dimensionless velocities ${\bf v}_a$,
corresponding to ${1\over c}{\bf v}_a$, $a=1,2$, in the above
references).

Similarly, for the gravitational case (following Sec.103 of
\cite{Fock}), one has to multiply the interaction term (i.e.,
$-{{\alpha_{{}_G}}\over{r}}$) in the Newtonian Lagrangian  with
the velocity dependent expression
\begin{eqnarray}
k_{{}_G}=k_{{}_{EM}}+{1\over 2}\left(3\left({\bf v}_1-{\bf
v}_2\right)^2-{{\alpha_{{}_G}}\over{r}}\right).
\label{kGr}
 \end{eqnarray}

We recall that a simple minded application of the above procedure
would give spurious first order effects that should be eliminated.

Noting that the CM velocities of the two particles are expressed
in terms of the  relative velocity ${\bf v}$ by
\begin{eqnarray}
{\bf v}_1\!=\!{{E_w}\over {E_1}}{\bf v},\,\,\,{\bf
v}_2\!=\!-{{E_w}\over {E_2}}{\bf v}\,\,\,\hbox{for}\,\,\,{\bf
v}_a\!=\!{{m_a}\over{E_a}}{\bf u}_a,\,\,\,{\bf
v}\!=\!{{m_w}\over{E_w}}{\bf u}
\label{vvu}
 \end{eqnarray}
we find
\begin{eqnarray}
k_{{}_{EM}}=1+{1\over 2}{{E_w^2}\over{E_1 E_2}}\left({\bf v}^2 +
v_r^2 \right),\\ k_{{}_G}=k_{{}_{EM}}+{1\over 2}\left(3{{w
E_w}\over{E_1 E_2}} {\bf
v}^2\!-\!{{\alpha_{{}_G}}\over{r}}\right).\nonumber
\label{kEMkG}
 \end{eqnarray}

In the above approximation one should replace the
energy dependent factors with their non-relativistic limits
\begin{eqnarray}
{{E_w^2}\over{E_1 E_2}}\to \nu,\,\,\,{{w E_w}\over{E_1 E_2}}\to
1\,\,\,\hbox{for}\,\,\,\varepsilon \to 0.
\label{nonr}
 \end{eqnarray}
We end up with the following simple expressions depending on the effective
particle 3-velocity ${\bf v}$ and position ${\bf r}$ only:

\begin{eqnarray}
k_{{}_{EM}}=1+{1\over 2}\nu\left({\bf v}^2 + v_r^2 \right),\\
k_{{}_G}=k_{{}_{EM}}+{1\over 2}\left(3 {\bf
v}^2\!-\!{{\alpha_{{}_G}}\over{r}}\right).\nonumber
\label{kEMkGeff}
 \end{eqnarray}

It is clear that these new terms will account for the differences
between  our result (\ref{dshift}) for $\delta \phi$ (obtained
neglecting retardation effects, i.e., under assumption
$k_{{}_G}\equiv 1$) and the result in \cite{OK89} \cite{SW93} (in
order $\rho^4$, $\rho^2\beta$). It is, however, challenging to
develop the corresponding calculational scheme in order to be able
to compare the third post-Newtonian approximation tackled in
\cite{DJS00}.

\vskip 1truecm

{\em Acknowledgments:} I.T. would like to thank Thibault Damour
for both stimulating the renewal of his interest in the
relativistic two-body problem and for his critical remarks on a
preliminary version of this paper. He acknowledges partial support
from the Bulgarian Council for Scientific Research under contract
F-828.

\end{document}